\documentclass[preprint,12pt]{elsarticle}
\usepackage{graphicx}
\usepackage{amssymb}
\usepackage{amsmath}
\usepackage{subfig}
\usepackage{multirow}
\usepackage{colortbl,color}
\usepackage{algorithm}
\usepackage[page,title,titletoc,header]{appendix}
\usepackage{multirow}
\usepackage{amsthm}
\usepackage{hyperref}
\usepackage{bm}
\biboptions{comma,sort&compress}
\journal{ }

\begin{document}

\begin{frontmatter}
\title{Appendix: Chapman-Enskog Expansion in the Lattice Boltzmann Method}
\author[KFUPM]{Jun~Li\footnote[1]{e-mail: lijun04@gmail.com}}
\address[KFUPM]{Center for Integrative Petroleum Research, \\ College of Petroleum Engineering and Geosciences, \\ King Fahd University of Petroleum $\&$ Minerals, Saudi Arabia}
\begin{abstract}
The Chapman-Enskog expansion was used in the lattice Boltzmann method (LBM) to derive a Navier-Stokes-like equation and a formula was obtained to correlate the LBM model parameters to the kinematic viscosity implicitly implemented in LBM simulations. The obtained correlation formula usually works as long as the model parameters are carefully selected to make the Mach number and Knudsen number small although the validity of Chapman-Enskog expansion that has a formal definition of time derivative without tangible mathematical sense is not recognized by many mathematicians.   
\end{abstract}
\begin{keyword}
lattice Boltzmann method \sep Chapman-Enskog expansion.
\end{keyword}
\end{frontmatter}

\section{Introduction}
We present the Chapman-Enskog expansion in LBM, which is based on the version in \cite{HeLuo1997} but has modifications that led to the general formula of strain rate tensor \cite{LiWang2010}.
\section{Basic algorithm of the ordinary LBM} 
The computational domain is uniformly discretized by using many grids with a constant distance $\Delta x$ at the $x,y,z$ directions and computational quantities are defined at those discrete grids. At each grid, we specify several lattice velocities $\vec e_\alpha$ indexed by $\alpha\in[0, Q-1]$ for $Q$ directions in total. $\vec e_\alpha$ either is static for $\alpha=0$ or transports particles from the current grid at $\vec x$ to its neighboring grids at $\vec x+\Delta t\vec e_\alpha$ during each time step $\Delta t$. The magnitude of $\vec e_\alpha$ depends on $c=\Delta x/\Delta t$. For example, in the two-dimensional D2Q9 model \cite{Qianetal1992}, $\vec e_0=(0,0)$ and $\omega_0=4/9$, $\vec e_\alpha=(\cos\theta_\alpha,\sin\theta_\alpha)c$ and $\theta_\alpha=(\alpha-1)\pi/2$ and $\omega_\alpha=1/9$ for $\alpha\in[1,4]$, $\vec e_\alpha=(\cos\theta_\alpha,\sin\theta_\alpha)\sqrt{2}c$ and $\theta_\alpha=(\alpha-5)\pi/2+\pi/4$ and $\omega_\alpha=1/36$ for $\alpha\in[5,8]$, where $\omega_\alpha$ is the weight factors. 

The basic unknown variable is the density distribution function $f_\alpha(\vec x, t)$, which is used to compute the flow velocity $\vec u$ and pressure $p$ at $(\vec x, t)$. Here, we discuss the algorithm only for the mass and momentum transportations in the absence of external force. The equilibrium distribution function is specified as follows: 
\begin{equation} \label{eq:feq}
\begin{aligned}
   f^{\rm eq}_\alpha=\rho\omega_\alpha[1+\dfrac{3}{c^2}\vec e_\alpha\cdot\vec u+\dfrac{9}{2c^4}(\vec e_\alpha\cdot\vec u)^2-\dfrac{3}{2c^2}\vec u\cdot\vec u], 
\end{aligned}
\end{equation}
where the density $\rho$ and flow velocity $\vec u$ are defined using $f_\alpha$ and will be introduced after Eq.~\eqref{eq:conservation laws}. 

The D2Q9 model as well as other lattice models satisfy the following important properties (\textit{note}: $\vec e_0=(0,0)$):
\begin{equation} \label{eq:E(n)}
\begin{aligned}
   E^{(n)}=\sum_{\alpha=0}^8\omega_\alpha e_{\alpha,i_1}e_{\alpha,i_2}\cdots e_{\alpha,i_n}, 
\end{aligned}
\end{equation}
where $i_1,\cdots,i_n\in[1,3]$ are indices for the $x,y,z$ directions and  
\begin{equation} \label{eq:detailedE(n)}
\begin{split}
   & E^{(0)}=\sum_{\alpha=0}^8\omega_\alpha=1, \\
   & E^{(2)}=\sum_{\alpha=0}^8\omega_\alpha e_{\alpha,i}e_{\alpha,j}=\dfrac{c^2}{3}\delta_{ij}, \\
   & E^{(4)}=\sum_{\alpha=0}^8\omega_\alpha e_{\alpha,i}e_{\alpha,j}e_{\alpha,k}e_{\alpha,l}=\dfrac{c^4}{9}(\delta_{ij}\delta_{kl}+\delta_{ik}\delta_{jl}+\delta_{il}\delta_{jk}), \\
   & E^{(2n+1)}=0, n=1,2,3,\cdots, \\
\end{split}
\end{equation}
where simpler indices $i,j,k,l\in[1,3]$ can be used for the $x,y,z$ directions. 

We use $\sum_\alpha$ as $\sum_{\alpha=0}^8$ for simplicity below. According to the above properties, we can compute the following low-order moments of $f^{\rm eq}_\alpha$: 
\begin{equation} \label{eq:moments of feq}
\begin{split}
   & \sum_\alpha f^{\rm eq}_\alpha=\rho, \\
   & \sum_\alpha e_{\alpha,i}f^{\rm eq}_\alpha=\rho u_i, \\
   & \sum_\alpha e_{\alpha,i}e_{\alpha,j}f^{\rm eq}_\alpha=\dfrac{c^2}{3}\rho\delta_{ij}+\rho u_iu_j, \\
   & \sum_\alpha e_{\alpha,i}e_{\alpha,j}e_{\alpha,k}f^{\rm eq}_\alpha=\dfrac{c^2}{3}\rho(\delta_{ij}u_k+\delta_{ik}u_j+\delta_{jk}u_i). \\
\end{split}
\end{equation}
As will be shown later, the properties in Eq.~\eqref{eq:moments of feq} is very important in the derivation of a Navier-Stokes-like equation from LBM. The general rule in constructing new lattice models $\vec e_\alpha, \omega_\alpha$ and $f^{\rm eq}_\alpha$  is to satisfy Eq.~\eqref{eq:moments of feq} and then a Navier-Stokes-like equation can always be recovered from LBM. 

The explicit updating algorithm of the only unknown $f_\alpha(\vec x,t)$ for each $\Delta t$ is a relaxation process toward $f^{\rm eq}_\alpha(\vec x,t)$:
\begin{equation} \label{eq:evolution falpha}
\begin{split}
   & f_\alpha(\vec x+\Delta t\vec e_\alpha,t+\Delta t)=f_\alpha(\vec x,t)+\dfrac{f^{\rm eq}_\alpha(\vec x,t)-f_\alpha(\vec x,t)}{\tau}, \\
\end{split}
\end{equation}
where $\tau$ is the normalized relaxation time. The relaxation process of Eq.~\eqref{eq:evolution falpha} should conserve mass and momentum as follows:  
\begin{equation} \label{eq:conservation laws}
\begin{split}
   & \sum_\alpha f^{\rm eq}_\alpha(\vec x,t)=\sum_\alpha f_\alpha(\vec x,t), \\
   & \sum_\alpha e_{\alpha,i}f^{\rm eq}_\alpha(\vec x,t)=\sum_\alpha e_{\alpha,i}f_\alpha(\vec x,t), \\
\end{split}
\end{equation}
which gives the definitions $\rho(\vec x,t)=\sum_\alpha f_\alpha(\vec x,t)$ and $u_i(\vec x,t)=\dfrac{1}{\rho}\sum_\alpha e_{\alpha,i}f_\alpha(\vec x,t)$ according to Eq.~\eqref{eq:moments of feq}. Now, Eq.~\eqref{eq:evolution falpha} is closed and has several parameters, including $\Delta x$, $\Delta t$ and $\tau$.   
\section{Derivation of Navier-Stokes-like equation} 
According to the Taylor expansion, we can rewrite Eq.~\eqref{eq:evolution falpha} into: 
\begin{equation} \label{eq:evolution falpha Taylor expansion}
\begin{split}
   & \sum_{n=1}^\infty\dfrac{\Delta t^n}{n!}D^n_tf_\alpha(\vec x,t)=\dfrac{f^{\rm eq}_\alpha(\vec x,t)-f_\alpha(\vec x,t)}{\tau}, \\
\end{split}
\end{equation}
where $D_t=(\partial_t+\vec e_\alpha\cdot\nabla)$. Now, the Chapman-Enskog expansion is introduced: 
\begin{equation} \label{eq:Chapman-Enskog expansion}
\begin{split}
   & f_\alpha=f^{(0)}_\alpha+\sum_{n=1}^\infty f^{(n)}_\alpha=f^{\rm eq}_\alpha+\sum_{n=1}^\infty f^{(n)}_\alpha, \\
   & \partial_t=\sum_{n=0}^\infty\partial_{t_n}, \\
\end{split}
\end{equation} 
where the expansion $\partial_t=\sum_{n=0}^\infty\partial_{t_n}$ of the time derivative is just a formal definition but not executable for any given analytical formula of $f_\alpha(\vec x,t)$ and thus this expansion has no tangible mathematical sense. 

Then,  terms in Eq.~\eqref{eq:evolution falpha Taylor expansion} can be sorted according to the order of magnitude and Eq.~\eqref{eq:evolution falpha Taylor expansion} can be replaced by a series of equations arranged into a consecutive order of magnitude: 
\begin{equation} \label{eq:splited eqs}
\begin{split}
   & \Delta t(\partial_{t_0}+\vec e_\alpha\cdot\nabla)f^{\rm eq}_\alpha=\dfrac{-1}{\tau}f^{(1)}_\alpha, \\
   & \Delta t(\partial_{t_0}+\vec e_\alpha\cdot\nabla)f^{(1)}_\alpha+\Delta t\partial_{t_1}f^{\rm eq}_\alpha+\dfrac{\Delta t^2}{2}(\partial_{t_0}+\vec e_\alpha\cdot\nabla)^2f^{\rm eq}_\alpha=\dfrac{-1}{\tau}f^{(2)}_\alpha. \\
   & \cdots
\end{split}
\end{equation} 
In order to make each $f^{(n)}_\alpha$ tractable in Eq.~\eqref{eq:splited eqs} and meanwhile the conservation rules of Eq.~\eqref{eq:conservation laws} is still satisfied, the following \textit{harsh} assumptions are used to replace Eq.~\eqref{eq:conservation laws} (\textit{note}: $f^{(0)}_\alpha=f^{\rm eq}_\alpha$ as assumed in Eq.~\eqref{eq:Chapman-Enskog expansion}): 
\begin{equation} \label{eq:harsh conservation laws}
\begin{split}
   & \sum_\alpha f^{(n)}_\alpha=0, \forall n\ne0, \\
   & \sum_\alpha e_{\alpha,i}f^{(n)}_\alpha=0, \forall n\ne0. \\
\end{split}
\end{equation}

By rewriting the second equation with the first one of Eq.~\eqref{eq:splited eqs} and using Eqs.~\eqref{eq:moments of feq} and \eqref{eq:harsh conservation laws}, the zero-order moments of the two equations of Eq.~\eqref{eq:splited eqs} (\textit{note}: check the definition in Eq.~\eqref{eq:moments of feq}) are:  
\begin{equation} \label{eq:zero order NS}
\begin{split}
   & \dfrac{\partial\rho}{\partial t_0}+\dfrac{\partial(\rho u_j)}{\partial x_j}=0, \\
   & \dfrac{\partial\rho}{\partial t_1}=0. \\
\end{split}
\end{equation}
Similarly, we can get the first-order moments of the two equations of Eq.~\eqref{eq:splited eqs}: 
\begin{equation} \label{eq:first order NS}
\begin{split}
   & \dfrac{\partial(\rho u_i)}{\partial t_0}+\dfrac{\partial}{\partial x_j}(\dfrac{c^2}{3}\rho\delta_{ij}+\rho u_iu_j)=0, \\
   & \dfrac{\partial(\rho u_i)}{\partial t_1}+(1-\dfrac{1}{2\tau})\dfrac{\partial}{\partial x_j}\sum_\alpha e_{\alpha,i}e_{\alpha,j}f^{(1)}_\alpha=0. \\
\end{split}
\end{equation}

Using Eqs.~\eqref{eq:moments of feq}, \eqref{eq:splited eqs}, \eqref{eq:zero order NS} and \eqref{eq:first order NS}, we have: 
\begin{equation} \label{eq:long a}
\begin{split}
   & \sum_\alpha e_{\alpha,i}e_{\alpha,j}f^{(1)}_\alpha=-\tau\Delta t\sum_\alpha e_{\alpha,i}e_{\alpha,j}(\partial_{t_0}+\vec e_\alpha\cdot\nabla)f^{\rm eq}_\alpha \\
   & =-\tau\Delta t[\dfrac{\partial}{\partial t_0}(\dfrac{c^2}{3}\rho\delta_{ij}+\rho u_iu_j)+\dfrac{\partial}{\partial x_k}\sum_\alpha e_{\alpha,i}e_{\alpha,j}e_{\alpha,k}f^{\rm eq}_\alpha] \\
   & =-\tau\Delta t[\dfrac{-c^2}{3}\delta_{ij}\dfrac{\partial}{\partial x_k}(\rho u_k)+\dfrac{\partial(\rho u_iu_j)}{\partial t_0}+\dfrac{\partial}{\partial x_k}\sum_\alpha e_{\alpha,i}e_{\alpha,j}e_{\alpha,k}f^{\rm eq}_\alpha], \\
\end{split}
\end{equation}
where 
\begin{equation} \label{eq:long b}
\begin{split}
   & \dfrac{\partial(\rho u_iu_j)}{\partial t_0}=u_i\dfrac{\partial(\rho u_j)}{\partial t_0}+u_j\dfrac{\partial(\rho u_i)}{\partial t_0}-u_iu_j\dfrac{\partial\rho}{\partial t_0} \\
   & =-u_i\dfrac{\partial}{\partial x_k}(\dfrac{c^2}{3}\rho\delta_{jk}+\rho u_ju_k)-u_j\dfrac{\partial}{\partial x_k}(\dfrac{c^2}{3}\rho\delta_{ik}+\rho u_iu_k)+u_iu_j\dfrac{\partial(\rho u_k)}{\partial x_k} \\
   & =-u_i\dfrac{c^2}{3}\dfrac{\partial\rho}{\partial x_j}-u_j\dfrac{c^2}{3}\dfrac{\partial\rho}{\partial x_i}-u_i\dfrac{\partial(\rho u_ju_k)}{\partial x_k}-u_j\dfrac{\partial(\rho u_iu_k)}{\partial x_k}+u_iu_j\dfrac{\partial(\rho u_k)}{\partial x_k} \\
   & =-u_i\dfrac{c^2}{3}\dfrac{\partial\rho}{\partial x_j}-u_j\dfrac{c^2}{3}\dfrac{\partial\rho}{\partial x_i}-\dfrac{\partial(\rho u_iu_ju_k)}{\partial x_k}
\end{split}
\end{equation}
and 
\begin{equation} \label{eq:long c}
\begin{split}
   & \dfrac{\partial}{\partial x_k}\sum_\alpha e_{\alpha,i}e_{\alpha,j}e_{\alpha,k}f^{\rm eq}_\alpha=\dfrac{\partial}{\partial x_k}[\dfrac{c^2}{3}\rho(\delta_{ij}u_k+\delta_{ik}u_j+\delta_{jk}u_i)] \\
   & =\dfrac{c^2}{3}\delta_{ij}\dfrac{\partial(\rho u_k)}{\partial x_k}+\dfrac{c^2}{3}\dfrac{\partial\rho}{\partial x_i}u_j+\dfrac{c^2}{3}\rho\dfrac{\partial u_j}{\partial x_i}+\dfrac{c^2}{3}\dfrac{\partial\rho}{\partial x_j}u_i+\dfrac{c^2}{3}\rho\dfrac{\partial u_i}{\partial x_j}.
\end{split}
\end{equation}
Substituting Eqs.~\eqref{eq:long c} and \eqref{eq:long b} into Eq.~\eqref{eq:long a}, we get: 
\begin{equation} \label{eq:strain rate}
\begin{split}
   & \sum_\alpha e_{\alpha,i}e_{\alpha,j}f^{(1)}_\alpha=-\tau\Delta t[\dfrac{c^2}{3}\rho(\dfrac{\partial u_j}{\partial x_i}+\dfrac{\partial u_i}{\partial x_j})-\dfrac{\partial}{\partial x_k}(\rho u_iu_ju_k)],
\end{split}
\end{equation}
which is used in \cite{LiWang2010} to estimate the strain rate tensor for applying large eddy simulations (LES) in LBM.  

Now, combining equations in Eq.~\eqref{eq:zero order NS} by using $\partial_t=\sum_{n=0}^\infty\partial_{t_n}\approx\partial_{t_0}+\partial_{t_1}$, we get: 
\begin{equation} \label{eq:combined zero order NS}
\begin{split}
   & \dfrac{\partial\rho}{\partial t}+\dfrac{\partial(\rho u_j)}{\partial x_j}=0. \\
\end{split}
\end{equation}
Combining equations in Eq.~\eqref{eq:first order NS} and using Eq.~\eqref{eq:strain rate}, we get: 
\begin{equation} \label{eq:combined first order NS}
\begin{split}
   \dfrac{\partial(\rho u_i)}{\partial t}+\dfrac{\partial(\rho u_iu_j)}{\partial x_j}= & -\dfrac{\partial}{\partial x_i}(\dfrac{c^2\rho}{3})+ \\
   & \dfrac{\partial}{\partial x_j}[(\tau-0.5)\Delta t(\dfrac{c^2}{3}\rho(\dfrac{\partial u_j}{\partial x_i}+\dfrac{\partial u_i}{\partial x_j})-\dfrac{\partial}{\partial x_k}(\rho u_iu_ju_k))]. \\
\end{split}
\end{equation}

The solutions of $\rho$ and $\vec u$ in LBM simulations satisfy Eqs.~\eqref{eq:combined zero order NS} and \eqref{eq:combined first order NS}, which are different from the standard incompressible Navier-Stokes equation. But, if we choose the model parameters (e.g., $\Delta x$, $\Delta t$ and $\tau$) carefully such that the magnitude of $\vec u$ is much smaller than $c/\sqrt{3}$ (i.e., the sound speed in LBM simulations), the relative variation of $\rho$ and $-\dfrac{\partial}{\partial x_k}(\rho u_iu_ju_k)$ of Eq.~\eqref{eq:combined first order NS} are negligible. Then, $-\dfrac{\partial}{\partial x_i}(\dfrac{c^2\rho}{3})$ and $\vec u$ correspond to $-\dfrac{\partial p}{\partial x_i}$ and $\vec u$ of the standard incompressible Navier-Stokes equation, where the kinematic viscosity is implemented in LBM according to Eq.~\eqref{eq:nu} as suggested by Eq.~\eqref{eq:combined first order NS}: 

\begin{equation} \label{eq:nu}
\begin{aligned}
   \nu=\dfrac{(\tau-0.5)\Delta tc^2}{3}. 
\end{aligned}
\end{equation}


\end{document}